\documentclass[journal]{IEEEtran}
\usepackage{amsmath}
\usepackage{graphicx}
\usepackage{epstopdf}
\usepackage{amssymb}
\usepackage{amsthm}
\usepackage{lipsum}
\usepackage{cite}
\usepackage{subfigure}
\usepackage{enumerate} 
\usepackage{hhline}
\usepackage{boldline}
\usepackage{tikz}
\usepackage{fancyhdr}
\usepackage{lettrine}
\usepackage{subfigure}
\usetikzlibrary{patterns}
\usetikzlibrary{shapes,arrows}
\usepackage{verbatim}

\usetikzlibrary{positioning}
\usetikzlibrary{shapes.geometric}
\usetikzlibrary{shapes.misc}
\usepackage{algorithmic}
\usepackage{booktabs}
\usepackage{multicol}
\usepackage[linesnumbered,ruled,vlined]{algorithm2e}
 \usepackage{siunitx}

\begin{document}
	
		\title{Indexed Multiple Access with Reconfigurable Intelligent Surfaces: The Reflection Tuning Potential}
	\author{Rohit Singh, \textit{Member IEEE}, Aryan Kaushik, \textit{Member, IEEE}, Wonjae Shin, \textit{Senior Member, IEEE,} George C. Alexandropoulos,  \textit{Senior Member, IEEE,} Mesut Toka, \textit{Member IEEE}, and Marco Di Renzo, \textit{Fellow, IEEE}	
 \thanks{R. Singh is with the Department of Electronics and Communication Engineering, Dr. B. R. Ambedkar National Institute of Technology, Jalandhar, 144027, India (email: rohits@nitj.ac.in)\\ $~~~~$A. Kaushik is with the School of Engineering and Informatics, University	of Sussex, BN1 9RH Brighton, UK. (e-mail: aryan.kaushik@sussex.ac.uk) \\ $~~~~$W. Shin and M. Toka are with the Department of Electrical and Computer Engineering, Ajou University, Suwon-si 16499, South Korea. W. Shin is also with the Department of Electrical Engineering, Princeton University, Princeton, NJ 08544
			USA (e-mail: \{wjshin, tokamesut\}@ajou.ac.kr).\\
			$~~~~$G. C. Alexandropoulos is with the Department of Informatics and
			Telecommunications, National and Kapodistrian University of Athens, 15784
			Athens, Greece and also with the Technology Innovation Institute, 9639
			Masdar City, Abu Dhabi, United Arab Emirates (e-mail: alexandg@di.uoa.gr).\\
		$~~~~$M. Di Renzo is with Paris-Saclay University, CNRS, CentraleSupelec,
		Laboratoire des Signaux et Systemes, 3 Rue Joliot-Curie, 91192 Gif-sur-Yvette, France. (e-mail: marco.di-renzo@universite-paris-saclay.fr).}}
\maketitle

\begin{abstract}
Indexed modulation (IM) is an evolving technique that has become popular due to its ability of parallel data communication over distinct combinations of transmission entities. In this article, we first provide a comprehensive survey of IM-enabled multiple access (MA) techniques, emphasizing the shortcomings of existing non-indexed MA schemes. Theoretical comparisons are presented to show how the notion of indexing eliminates the limitations of non-indexed solutions. We also discuss the benefits that the utilization of a reconfigurable intelligent surface (RIS) can offer when deployed as an indexing entity. In particular, we propose an RIS-indexed multiple access (RIMA) transmission scheme that utilizes dynamic phase tuning to embed multi-user information over a single carrier. The performance of the proposed RIMA is assessed in light of simulation results that confirm its performance gains. The article further includes a list of relevant open technical issues and research directions.
\end{abstract}

\begin{IEEEkeywords}
	Reconfigurable intelligent surface, phase tuning, non-orthogonal multiple access, multi-user network.
\end{IEEEkeywords}

\section{Introduction}
Multiple access (MA) schemes have been a key enabler for the paradigm shift in the previous wireless generations \cite{ma}. To fulfill the ever-increasing data rate demand, each subsequent wireless generation (G) fuels research on spectral efficient MA mechanisms via spectral reuse. Spectral reuse can be achieved in several ways, including non-orthogonal multiple access (NOMA),  spatial division multiple access (SDMA), etc. However, the advent of indexed modulation (IM) has gained significant popularity due to its outstanding features. Specifically, data indexing via IM adds another degree of freedom (DoF) in the form of distinct combinations of transmitter antennas, radiation patterns, frequencies of operation, or their combinations. Basically, IM enables to map the transmission bits by tuning the on/off status of their transmission entities including subcarriers, transmit antennas, etc. Another parallel research is being conducted on programmable wireless environment configurations using reconfigurable intelligent surface (RIS). RIS is a programmable structure consisting of numerous electrically tunable elements that are capable of tuning the phases of the received signal. To do this, the phase shift at each RIS element can be electrically controlled using a smart controller supported by a transmission unit. Interestingly, the phase-controlling ability of RIS makes it favorable for data indexing, where different combinations of RIS elements can also be utilized for information mapping. Unlike antenna indexing, RIS-enabled data indexing can provide the following benefits \cite{ran}: \textit{a)} better energy efficiency against antenna indexing since RISs are passive in nature, \textit{b)}  infinitely large number of data combinations as each of RIS unit contains numerous elements, \textit{c)}  better feasibility as RIS can be easily deployed over any planar surfaces, etc. It can be inferred from the above discussion that data-indexing requirements and RIS features complement each other. Combining the benefits of RIS, this work brings together several innovative aspects of IM-enabled transmission and RIS-added benefits. Specifically, we propose the utilization of RIS-enabled data indexing via phase tuning along with a  comprehensive framework of RIS-enabled indexing. Moreover, this work pinpoints challenges, use cases, and open issues in RIS added transmissions, and the performance of the proposed RIS-indexed multiple access (RIMA) is discussed in light of simulation results.

\section{Recent Indexed and Non-indexed MA Solutions}
The notion of IM is introduced to meet the ever-increasing high throughput demand and energy efficiency. IM is an innovative modulation technique that utilizes activation states of transmission resource(s) for information encoding. These resources are either physical (i.e., time slot, subcarrier, antenna, etc) or  virtual (i.e., wireless channels) entities. This section presents the intuition of IM along with its classification.

\subsection{Indexed Transmission and Forms}
The principle of IM is based on choosing a combination from already tagged entities that naturally reveals the mapped information at the receiver, mainly using different entities or their combinations. For example, the selection of $k$ antennas (from the available $N$ antennas at the transmission station) forms  $m= {N \choose k}$ combinations, each referring to a distinct constellation point, often referred to as code-book. Accordingly, each transmission naturally carries $\lfloor \log_2 m\rfloor$ binary bits. 

Based on the nature of transmission characteristics, IM can be classified as:

\begin{table*}
	\caption{Comparison of different multiple access schemes}
	\begin{center}
		\centering
		
		\begin{tabular}{|p{0.12\linewidth}|p{0.16\linewidth}|p{0.18\linewidth}|p{0.18\linewidth}|p{0.2\linewidth}|}
			\hlineB{3}
			\textbf{Multiple Access} & \textbf{NOMA} & \textbf{SDMA} \cite{sdma}& \textbf{RSMA} \cite{rsma}& \textbf{Proposed RIMA}\\
			\hline
			\hline
			\textbf{Principle} & Fully decode interference & Treat interference as noise & Partial decoding of interference & No interference decoding required\\
			\hline
			\textbf{Decoder} & SIC performed at receiver & Interference as noise & SIC performed at receiver & SIC not required\\
			\hline
			\textbf{Network Load} & More suitable for overload network & Suitable for under-load network & Any network & Equally favourable for loaded network\\
			\hline 
			\textbf{Experienced SINR ($\Gamma$)} & $\Gamma_w = \frac{P_w ||\textbf{g}_w||^2}{P_s ||\textbf{g}_w||^2+ P_n}$, (for weak user)  & $\Gamma_w = \frac{P_t  ||{\textbf{p}_w} \textbf{g}_w||^2}{P_t  ||{\textbf{p}_s} \textbf{g}_w||^2+ P_n}$ & $\Gamma_w = \frac{P_t  ||{\textbf{p}_w} \textbf{g}_w||^2}{P_t  ||{\textbf{p}_s} \textbf{g}_w||^2+ P_n}$, (for common stream)  & $\Gamma_w = \frac{P_t  || \textbf{g}_w||^2}{P_n}$\\
			\hline \multicolumn{5}{|l|}{{Here, $P_t$ denotes transmitting power, subscripts ``$s$'' and ``$w$'' respectively denote strong and weak user of} NOMA and one of the group users for}\\
			\multicolumn{5}{|l|}{other schemes, $\textbf{g}_k$ and $\textbf{p}_k$ respectively are the channel coefficient and pre-coding vector for k$^{th}$ user, and $P_n$ is the noise power.}\\
			\hlineB{3}
		\end{tabular}
		\label{tab:tab1}
	\end{center}
\end{table*}

\textbf{Frequency Domain Index Modulation (FD-IM): }  FD-IM is possible when multiple sub-bands are available that enables to use various combinations of available frequency bands. Specifically, the presence of multiple antennas (i.e., placed sufficiently apart to experience uncorrelated channels) enables to orthogonalize the different code-books. Some of the well-known FD-IM schemes include subcarrier-IM, orthogonal frequency division multiplexing-IM (OFDM-IM), etc., that extend the notion of data indexing to different available subcarriers. It is evident from the recent works including \cite{wen} that FD-IM outperforms the plain OFDM.   

\textbf{Antenna-code Domain Index Modulation (AD-IM):}
AD-IM requires multiple transmitter antennas where information mapping is done by selecting a combination of a few antennas from the available ones \cite{majorcom}. The transmission antenna is active if it radiates the transmission power. Further, the ON/OFF state of transmission antennas naturally reveals the information at the receiver. For instance, in the above-mentioned example, only $k$ antennas activate at a given time while $N-k$ remain inactive.  Accordingly, the indexing of $k$ active antennas reveals the transmitted bits.

\textbf{Time-space Domain Index Modulation (TD-IM):} TD-IM uses the mapping of transmitted bits across multiple time slots in the diversified multiple input multiple output (MIMO) systems. Specifically, TD-IM converges from the classical approach of space-time matrix designs,  shifting toward the exploration of the space-time
for information mapping.

\subsection{Evolving Non-indexed Solutions}
Apart from the above-mentioned schemes, the notion of channel estimation and precoder based transmission has evolved the following MA schemes: 

\textbf{NOMA:} This enables to serve multiple users over the same time-frequency resource utilizing the notion of power split among distinct users' DoF \cite{noma}. High power allocation to the weak user with lower channel gain boosts up the intended signal power at the receiver. Moreover, the  user with a stronger channel first estimates the information of the weak user and then performs successive interference cancellation (SIC) for the intended signal separation. 

\textbf{SDMA:} SDMA is implemented in MIMO systems that utilize uncorrelated channels as another DoF allowing multiple simultaneous transmissions over the same time-frequency resource \cite{sdma}. Nevertheless, the benefit comes at the cost of rigorous channel estimation and precoding.  Accordingly, SDMA frameworks usually involve a large computational complexity.

\textbf{Rate Splitting Multiple Access (RSMA):} RSMA relies on partially decoding the interference and treating the other part of interference as noise \cite{rsma}. Doing so, RSMA combines the benefits of both the power domain NOMA and the SDMA. Moreover, the splitting power adjusts the level of interference and provides a balanced utilization of the available slot, spectrum, and power.

However, the mentioned MA schemes possess their own advantages and limitations including; \textit{a)} treating interference as noise affects users’ performance, \textit{b)} SIC is performed at the strong user that creates additional burden at the user-end. SIC involves additional processing time and causes error propagation because of imperfect SIC under the consideration of more than two users.  In more detail, the features and limitations of evolving MA schemes are highlighted in Table \ref{tab:tab1}, and are summarized below;
\begin{itemize}
	\item \textit{Principle:} Sharing of the same resource among multiple users leads to inter-user interference which is usually considered as the noise at the weak user and hence affects the signal-to-interference-noise Ratio (SINR).      
	\item \textit{Decoding:} Often, the strong user undergoes the SIC process to mitigate interference from the weak user. Accordingly, the signal of the weak user is detected and subtracted from the received signal by the strong user.
	\item \textit{Signal Strength:} Though NOMA, SDMA, and RSMA are claimed to outperform the conventional orthogonal transmission schemes, the users most often experience reduced SINR and have to undergo complex signal processing.  
\end{itemize}

\subsection{Theoretical Comparisons}
Given the aforementioned issues encountered in the non-indexed MA schemes, this section summarises the benefits of indexed MA against the non-indexed ones. Specifically, the benefits of IM transmission are summarized as:

\textbf{No physical transmission:} IM performs information-based indexing that appears at the mapped location of the constellation that reveals the desired information at the receiver without a physical transmission.
	
\textbf{Power split is not required:} IM information is embedded in the primary signal transmission without splitting powers among users, thus it prevents the reduction in the experienced signal strength, which makes it more suitable for diverse applications on a single platform.

\textbf{Getting rid of SIC:} Unlike non-indexed schemes including NOMA and RSMA, indexed transmission carries single information whereas the tagged information is superimposed on the indexed entities. Therefore, the data is prevented from other interfering sources that naturally eliminate the need for SIC. Accordingly, the user can now enjoy SIC-free service with dual transmission benefits.  

\section{Combining Benefits: RIS Enabled IM}
This section summarises how the benefits of IM and RIS can be combined together via RIS-enabled data encoding. 

\subsection{RIS Based Indexing}
Already, recent works have initiated research on RIS-based data indexing. For example, RIS-assisted NOMA with two users has been proposed in \cite{rnoma}, where both users experience RIS coordination in downlink transmission. Moreover, some early attempts on RIS indexed modulation include RIS-based transmission \cite{r2} which mainly amalgamates RIS configuration and index modulation at the receiver antenna indices. On the other hand, Lin et. al. \cite{r3} utilized a reflection pattern of distinct RIS combinations for data modulation to achieve passive beamforming and information transfer. Later, Li. et. al. \cite{r1} implemented RIS-based modulation via information-theoretic element selection. However, the aforementioned works \cite{r2,r3,r1} only considered the configuration ability of RIS network to enhance strength and considerably neglected the fact that distinct RIS elements and phase shift configurations can be used for signal encoding. One step ahead, in this work (i.e., in Section \ref{s4}) we demonstrate that different combinations of RIS elements along with its phase-shifting ability can be used to encode distinct codewords in a constellation, owing to the uncorrelated channels (i.e., when inter-group distance is sufficiently larger than operating
wavelength).

\subsection{Choice of Indexing Entity}
The previous section lists several benefits of RIS-enabled transmission. Moreover, this section shows that RIS is even more favorable for data indexing where element-wise data encoding could be utilized for indexing. Different from AD-IM/TD-IM, RIS indexing intends to provide the following advantages \cite{rui}:    

\textbf{Cost and Energy Efficient:} From an indexing perspective, each RIS element can play the same role a typical antenna does. Nevertheless, multiple neighboring RIS elements are often grouped together to avoid correlation considering a single channel coefficient of the block. Still, the cost of each RIS group is expected to be several times lesser than that of the transmitter antenna. Moreover, RIS is passive in nature and hence becomes more favorable from an energy perspective too.      

\textbf{Ease of Implementation:} Unlike antennas that are often installed at the transceiver ends, RISs are flexible and can be installed anywhere in the transmission medium. Moreover, the planar geometry makes them easier to deploy on the walls, ceilings, or even the rooftops of moving vehicles.  This feature makes RIS a key technology of the forthcoming wireless generation.

\textbf{Boundless Combinations and Propagation Benefits:} RIS units can consist of thousands of elements $N$ that bring thousands of indexed constellation points $m\!=\!{N \choose k}$, i.e., grouping neighbourhood elements to prevent correlation. Infinitely available constellation points \textit{m} naturally increases the number of bits per transmission. Moreover, the presence of passive RIS provides propagation benefits as a complement.

\section{RIS Phase Tuning for Data Indexing}\label{s4}
Considering the propagation/indexing benefits of intelligent surfaces and  features of IM, we propose a novel RIS phase tuning method for data embedding that brings together the best of both worlds. 

     \begin{figure}
	\centering
\includegraphics[width=0.9\linewidth]{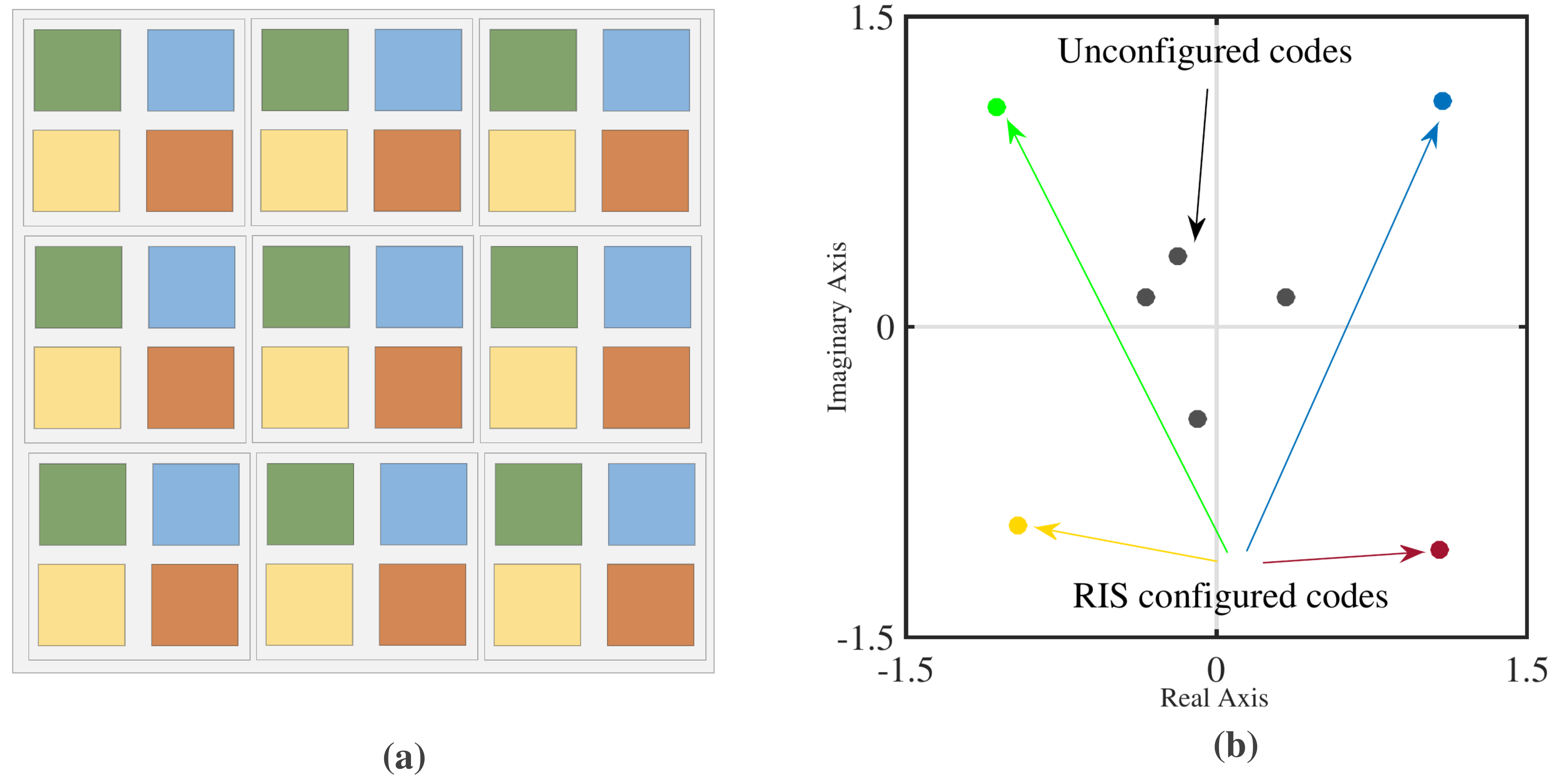}
\caption{An illustration of a) elemental-based phase tuning, b) and tuned constellation simulated via MATLAB. }
\label{fig:color}
\end{figure}

\subsection{Intuition}
The notion of RIS phase tuning, i.e., RIMA, is inspired by the ability of the RIS unit to provide the required phase shift in the constellation. RISs are capable of aligning the resultant phasors at the desired angle via appropriate phase tuning of the participating elements. In RIMA, the phase tuning is done on the basis of the constellation position of the transmitting information. For example, an illustration of phase tuned constellation is given in Fig. \ref{fig:color} where elements with the same color align the resultant sum at the desired angle (i.e., shown by the respective color in the constellation diagram). Accordingly, the elements with the same color are turned ON as per the transmitted information, and the rest of the elements operates in the absorption/coordination mode.

\begin{figure*}
	\centering
	\includegraphics[width=0.75\linewidth]{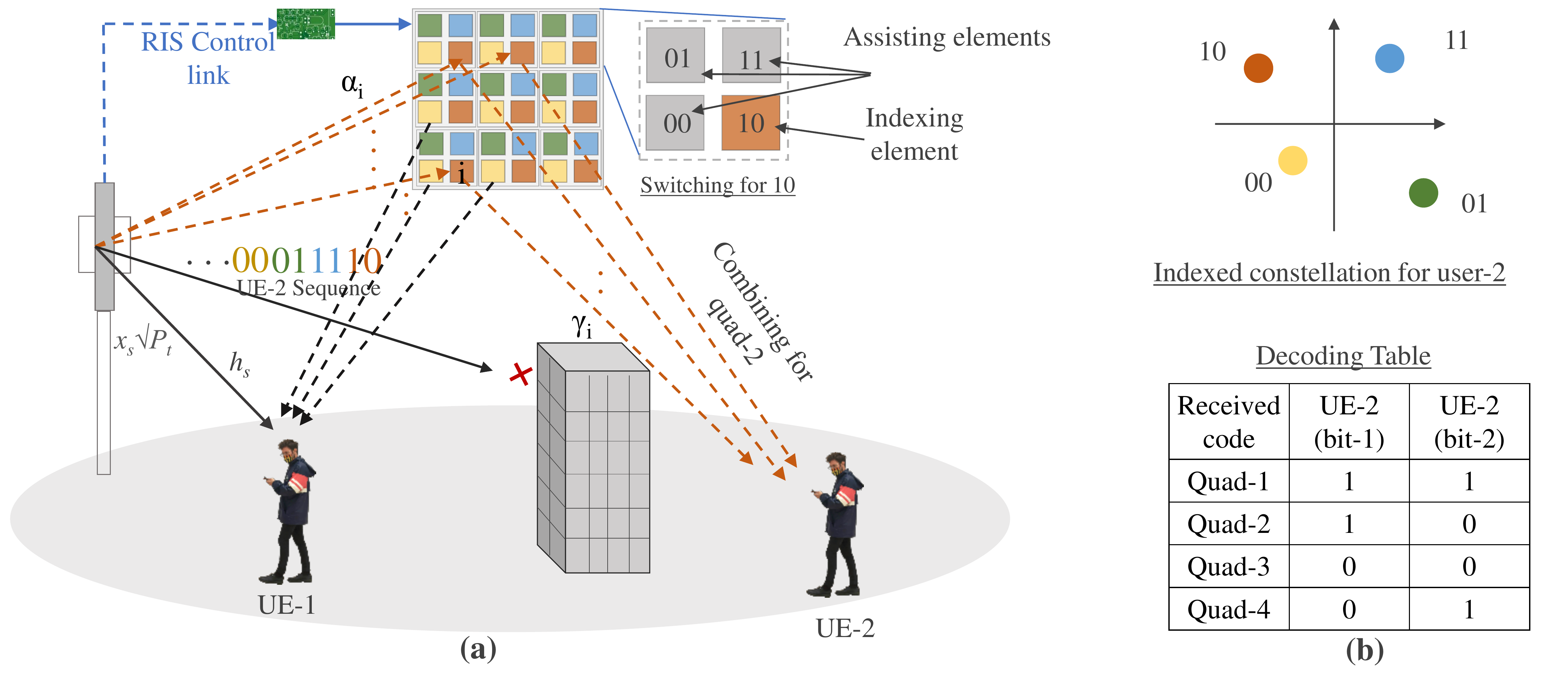}
	\caption{An illustration of RIMA transmission.}
	\label{fig:method}
\end{figure*}

\subsection{Methodology}
The section describes the methodology of RIMA transmission which is based on the phase-controlling ability of RIS units to tune the  constellation points on the basis of transmitted information. The methodology is illustrated in Fig. \ref{fig:method}, where phases of passive RIS elements are tuned via shared information, and controlling actions are taken by the base station (BS). For ease of understanding, a single transmitter and receiver antenna scenario is considered. Similar to the existing works \cite{rui}, Channel State Information (CSI) is assumed to be known at the BS and hence primary and secondary users are selected based on the instantaneous channel condition. Inspired by NOMA transmission, RIMA is intended to serve the strong user via transmitting its information (i.e., considered as the primary user in the given example) and the weak user via RIMA indexing (i.e., considered as the secondary user in the given example). In other words, the primary user, i.e., UE-1 with good channel condition in the figure, is served via the conventional method through precoding, i.e., where participating RIS elements assist UE-1. Whereas to serve the secondary user, no additional signal is transmitted, rather a pool of the same color indexing RIS elements carry the data based on the user's information. On the other hand, the rest of the elements assist the primary user in a manner that they appear inactive to the secondary users via creating null towards them. Accordingly, BS adapts RIS phases as per the channel condition plus instantaneous data, and hence the RIS phases now carry the transmitted information at the receiver side. The role of participating elements (i.e., indexing or assisting) switches as per the secondary user's data. Hence, extracting the phase of the received codeword encodes the transmitted information to the secondary users. Fig. \ref{fig:method} illustrates a two-user scenario (i.e., one `primary' and one `secondary'), nevertheless, RIMA can be easily extended to support multiple primary and secondary users over the same resource without power distribution. 

As depicted in Fig. \ref{fig:method}, the total $N_E$ RIS elements are mapped to one of the constellation codes, i.e., $b$ simultaneous bits forms $2^b$ codes, and hence the number of constellation points $N_G$ are $N_G=2^b=4$ for $b=2$. Accordingly, $N_R=\frac{N_E}{N_G}$ indexing elements participate in the instantaneous transmission for the secondary user (since the rest of the assisting elements effectively create null towards the secondary user).  Accordingly, let $h_s$ and $h_w$ be channel coefficients of links BS-(UE-1) and BS-(UE-2), respectively. When $|h_w|^2 < |h_s|^2$ is true, UE-1 and UE-2 are considered as primary and secondary users (the strong and weak users), respectively. Therefore unlike NOMA, RIMA enables to serve the strong user by transmitting the appropriate information $x_s$ with total available power $P_t$ (without splitting). On the other hand, no dedicated transmission takes place for the weak user. Instead, the same energy in the form of primary signal $\sqrt{P_t} x_s$ is released via tuning appropriate RIS elements based on the weak user's information as depicted in Fig. \ref{fig:method}.

For the weak user, the phases are controlled such that the resultant phasors are oriented at the required angle, i.e., based on the user's information. Considering RIS elements are tuned to instantaneous stream `10', then the received signal would appear in the second quadrant of the constellation. Accordingly, by detecting the relative angle, the transmitted message can be decoded at the receiver. In this manner, both users experience interference-free transmission without any SIC process. Also, the proposed RIMA scheme leads to the maximum received power since a power split is not required.

\subsection{Numerical Results}
This section illustrates numerical results for the proposed RIS-aided framework RIMA via MATLAB. For the sake of simplicity, we consider that the BS has the perfect CSI corresponding to the users. Since each RIS group leads to a different constellation point, the effectiveness of the proposed scheme is measured in terms of its achievable Bit Error Rate (BER) performance. A dual-phase transmission scenario has been simulated under independent and identically distributed Rayleigh fading channels, where the message stream of secondary user $u_w$ is transmitted via two RIS combinations each with distinct phases, i.e., $0$ and $\pi$. Moreover, the performance has been compared with the following two baseline schemes; \textit{i)} conventional NOMA scheme with two users \cite{ber}, where each of them shares a fraction of transmitted power, usually more power is assigned to the weak user. Accordingly, both the users experience interference that degrades their SINR performance, ii) RIS assisted NOMA with two users \cite{rnoma}, where both the users experience RIS coordination in downlink transmission. The target user $u_w$ experiences the worse channel condition, i.e., $|h_w|^2<|h_s|^2$, which confirms the same baselines for NOMA and RIMA. Accordingly, the target user is the weak/far user in the NOMA transmission scenario, and hence it does not apply the SIC technique at the receiver.  
 
 \begin{figure}[t]
	\centering
	\includegraphics[width=0.7\linewidth]{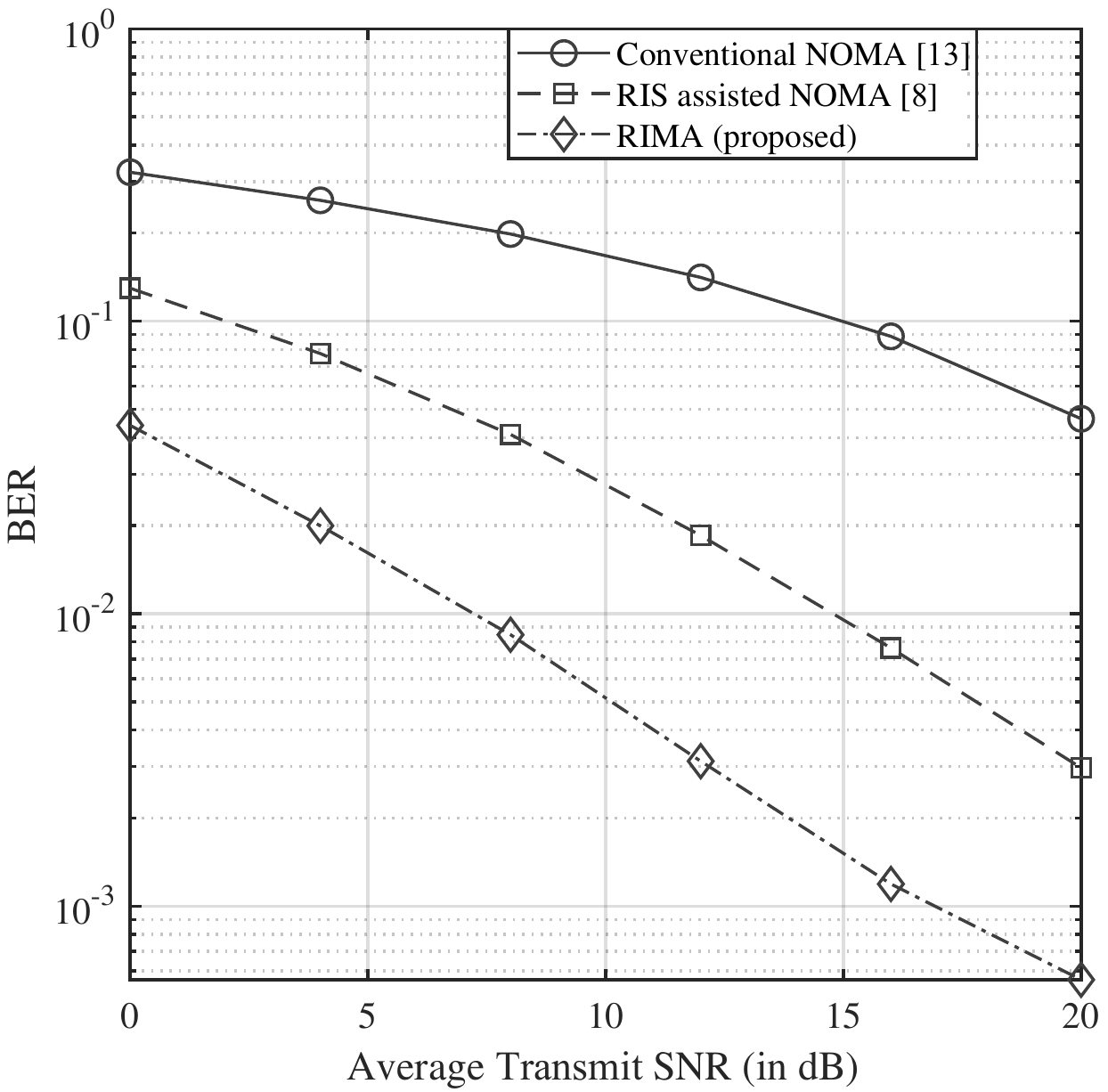}
	\caption{BER performance comparisons of different transmission schemes for BPSK.}
	\label{fig:BER_comp}
\end{figure}

\begin{figure}[t]
	\centering
	\includegraphics[width=0.7\linewidth]{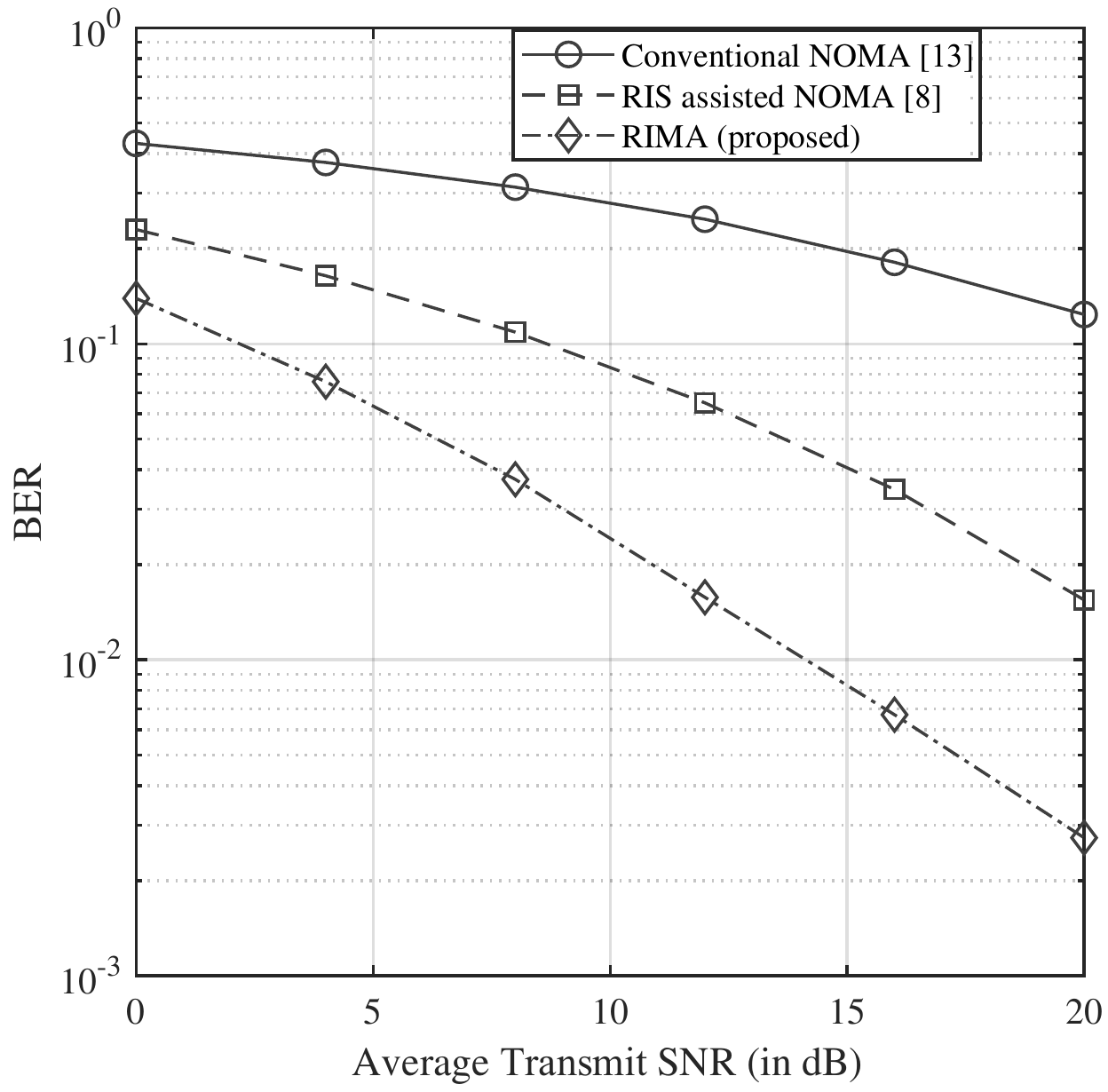}
	\caption{BER performance comparisons of different transmission schemes for 8-PSK. }
	\label{fig:BER_8PSK}
\end{figure}

Figs. \ref{fig:BER_comp} and \ref{fig:BER_8PSK} respectively illustrate the achievable BER performance of the weak user versus average transmit SNR in terms of binary phase shift keying (BPSK) and 8-PSK modulations for different schemes. $N_R = 12$ has been considered for RIMA and $\mu = 0.7$, i.e., a fraction of the total assigned to weak user, is considered for NOMA users. Furthermore, all the curves have been averaged over $10^5$ iterations to measure the mean performance. In general, the BER decreases with the increase in SNR for all the mentioned schemes. This is because the desired signal strength becomes more dominant against the noise as shown above. The curves demonstrate that even with a small number of elements, i.e., $N_R = 12$,  RIMA outperforms the baseline schemes by a significant margin. Specifically, the performance enhancement is achieved via the following groundbreaking features of RIMA; \textit{a)} the NOMA users cannot enjoy the full potential of transmitted power since a fraction of power is split to serve another user. Unlike NOMA, the information of the weak user carries over the primary signal and hence RIMA user does not experience a power split, \textit{b)} in RIMA, single transmission carries information for both the signal. Hence, the desired signal does not interfere with  the secondary user. Moreover, the BER performance for 8-PSK modulation is relatively poor which is obvious due to the reduction in inter-code space against the BPSK transmission. Nevertheless, it is also noteworthy that RIMA outperforms the baseline schemes in both  cases which shows that the proposed RIMA maintains its effectiveness even in closely packed constellation scenarios. It is worth noting that the indexed solution makes use of partial entities, e.g., only $N_R$ RIS elements are used to serve the secondary user while $N_E-N_R$ elements work in assisting mode in each transmission that virtually sets a trade-off between indexing users/bits and participating elements.

\begin{figure*}[!ht]
	\centering
	\includegraphics[width=0.75\textwidth]{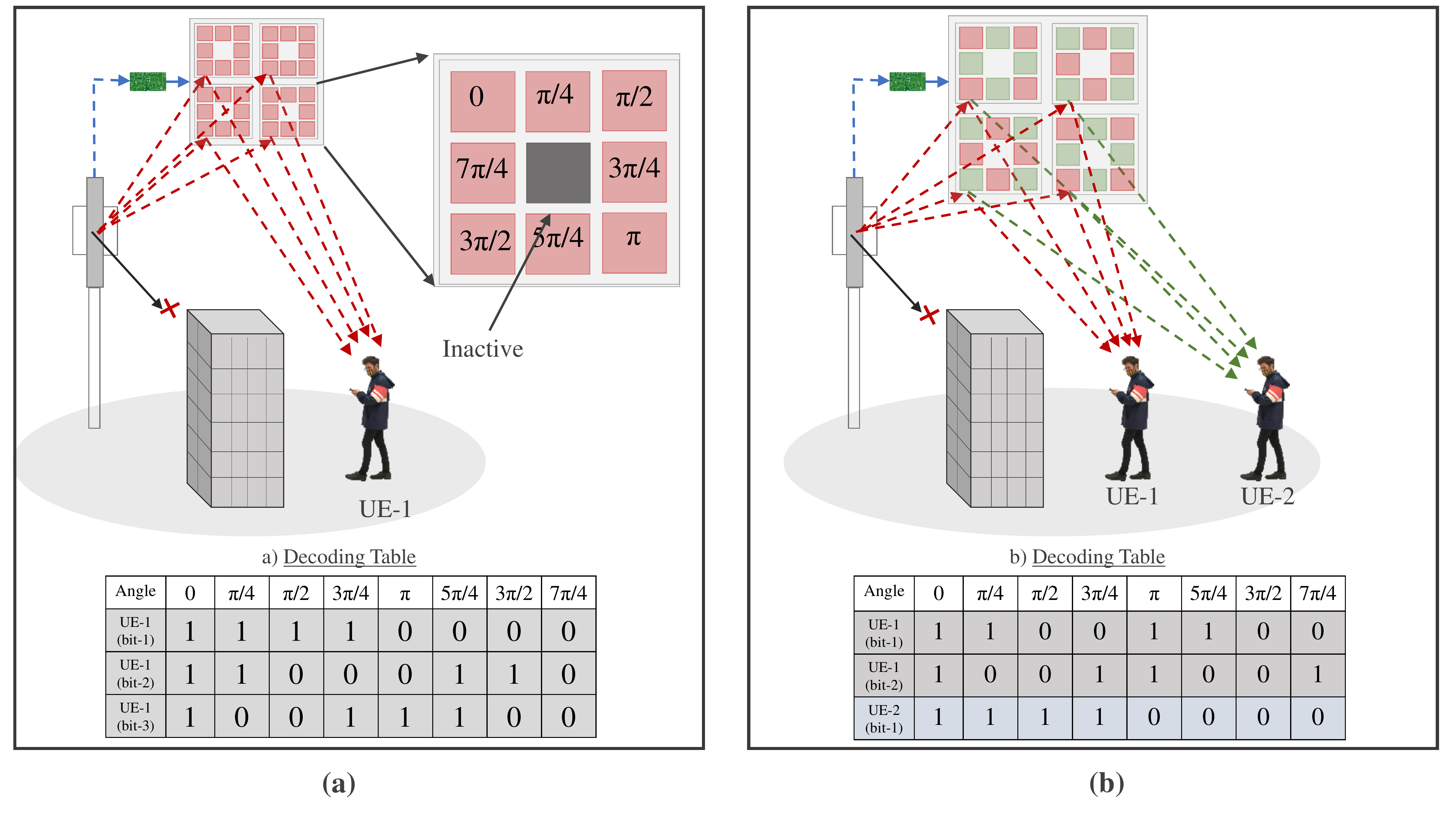}
	\caption{An illustration of multi-rate/multi-user adoption; a) 8-PSK indexing for a single user, b) multi-user RIMA transmission.} 
	\label{fig:two_user}
\end{figure*}

\section{Interplay with Future Communication Networks}
The forthcoming wireless world is going to support variety of applications with distinct quality of service (QoS) requirement. This subsection highlights another novel aspect of the proposed scheme by showing that RIMA is flexible and can be easily applied to multi-rate and/or multi-user scenarios. Fig. \ref{fig:two_user} illustrates a multi-user implementation of RIMA where RIS elements are arranged to provide eight combinations of distinct phases which can convey three bits simultaneously. Moreover, these programming input bits may belong to the same user or distinct users depending on the mode of transmission. For example, Fig. \ref{fig:two_user}(a) depicts a scenario where all three bits belong to the same user, where all the participating elements are assigned the same user, and hence the decoding Table resembles 8-PSK for UE-1. On the other hand, Fig. \ref{fig:two_user}(b) shows a scenario where RIS is operated under 8-PSK mode and the three-bit carrying capacity is shared between two users via distinct sent of indexing elements (i.e., denoted by different colors), each creating nulls towards unintended users. For instance, UE-1 is assumed to have a high QoS requirement and hence it is assigned two former bits while the last bit belongs to UE-2.

\section{Open Issues and Research Directions}
RIS-based data indexing proves to be an efficient solution from cost, energy, and implementation perspectives. Despite these traits, there are many issues that need to be carefully considered.

\subsection{Channel Estimation}
Similar to other RIS-added frameworks, the performance of RIMA is pillared on the accuracy of channel estimation. The presence of a large number of elements significantly increases the RIS training overhead. Specifically,  RISs are often trained either by using RF chains or by element-wise passive reflections. However, both the methods possess the following limitations: \textit{a)} element-wise passive training via turning ON the single element in each respective slot leads to a large training overhead, and \textit{b)} inclusion of active RF chains for parallel estimation increases the energy consumption and implementation cost. Another challenge is the cascading nature of the channel between the transmission point and the user end. Nevertheless, the channel between RIS and transmission unit lasts longer, i.e., owing to the static positions. Moreover, the research on low complexity RIS training is underway and some promising solutions are expected soon \cite{Jian2022, 9769689}. 

\subsection{Central Processing Requisites}
RISs are passive in nature and hence all the processing and controlling actions are performed at the central unit. Effectively, the central unit not only sends the required data but also performs pre-processing prior to the information transmission. Specifically, the central unit is responsible for the following key operations: 

\textbf{Pilot Collection and CSI Estimation:} Irrespective of the training method (i.e., RF-chain based or element-wise), the central unit has to collect a huge amount of training data required for the channel acquisition by each participating RIS element. Further, data acquisition is followed by rigorous signal processing to  estimate the channel coefficients. Moreover, the central processor has to periodically update the channel coefficients due to the varying nature of the channel. Due to a large number of elements, RIS's training and controlling processes exert a lot of overhead on the central units. Load sharing and edge computing can be potential solutions towards the mentioned issues.  

\textbf{Precoding and RIS Phase Controlling:} Once training data is obtained and channels are estimated, these coefficients are used to design a precoder that ensures the following: better signal strength at the receiver, interference avoidance towards the unintended users, or a combination of both, etc. However, the procedure for the precoder design involves complex computation that non-linearly increases with the participating transceiver points. Moreover, the central unit is also responsible for controlling the amount of phase shift at each participating RIS element. The inclusion of an active RF-chain undoubtedly releases central overhead but it affects the cost and energy consumption of the system. Recent works are focusing on RISs grouping with one central RF unit, however, the trade-off between overhead reduction and energy/cost increment still remains unstudied.

\textbf{Data Transmission and Embedding:} This phase comes after pre-processing (i.e., CSI Estimation and Precoding) that mainly involves the transmission of precoded information towards the intended users. Moreover, an IM-based framework gives additional responsibility to the central unit in the form of one-to-one information embedding to each combination of the involved entity. In this manner, the central unit has to take care of the primary transmission along with the secondary data embedding that exerts additional load on the central processor.

\subsection{Inter-element Correlation \& Bit-level Synchronization}
Apart from the above-mentioned facts, there are some questions that remain unanswered, for example: \textit{a)} RIS is well known for its ability to alter environmental impacts via its phase-controlling ability. Nevertheless, the performance gain is highly dependent on the independent channel coefficients at each RIS element, though the inter-element correlation may hinder the RIS-added gain. Since most of the work considers uncorrelated coefficients, the impact of inter-elements correlation still remains unstudied, \textit{b)} another challenge is to perform a bit-level synchronization at the transceiver end. Assuring bit-level synchronization is crucial since it is the backbone of data indexing. 

\section{Conclusive Remarks}
In this work, we propose a novel RIS indexing scheme for non-orthogonal transmission to overcome the limitations of the existing multiple access schemes. Specifically, this work highlights the shortcomings of existing non-indexed MA schemes and provides a theoretical comparison to show that indexing-based MA eliminates the limitations of non-indexed schemes. Also, it is shown that RIS provides several benefits as compared to antenna indexing. Furthermore, a RIS-Indexed Multiple Access scheme has been proposed along with the simulation results that confirm the performance gain achieved by RIMA. Finally, this work summarises open issues and research directions that shed light on the upcoming research on RIS-enabled MA.

\bibliographystyle{IEEEtran}

\bibliography{IEEEabrv,BibRef}

\end{document}